\documentstyle[preprint,aps,epsf,floats]{revtex}

\def\lb{\label}

\begin{document}

\draft

\title{A Beam Splitter for Guided Atoms on an Atom Chip}

\author{Donatella Cassettari$^1$, Bj\"orn
Hessmo$^{1,2}$, Ron Folman$^1$, Thomas Maier$^3$, J\"org
Schmiedmayer$^1$}

\address{$^1$Institute f\"ur Experimentalphysik,
Universit\"at Innsbruck, A-6020 Innsbruck,  Austria\\
$^2$Department of Quantum Chemistry, Uppsala University, S-75120
Uppsala, Sweden\\ $^3$Institute f\"ur Festk\"orperelektronik,
Floragasse 7, A-1040 Wien, Austria}

\date{\today}

\maketitle

\begin{abstract}
We have designed and experimentally studied a simple beam splitter
for atoms guided on an {\em Atom Chip}, using a current carrying
Y-shaped wire and a bias magnetic field. This beam splitter and
other similar designs can be used to build atom optical elements
on the mesoscopic scale, and integrate them in matterwave quantum
circuits.
\end{abstract}


\vskip1pc

%]

Beam splitters are key elements in optics and its applications. In
atom optics \cite{AtomOptics} beam splitters were, up to now, only
demonstrated for atoms moving in free space, interacting either
with periodic potentials (spatial and temporal), or semi
transparent mirrors \cite{ertmer}. On the other hand, guiding of
atoms has attracted much attention in recent years and different
guides have been realized using magnetic potentials
\cite{Schm92,WireGuide,mesoscop,ChipGuide_large,Hindsguide,ron},
hollow fibers \cite{HollowFibers}, and light potentials
\cite{donught,channeling}.

\begin{figure}
    \begin{center}\hspace{0mm}\mbox{\input epsf
\epsfxsize\columnwidth\epsfbox{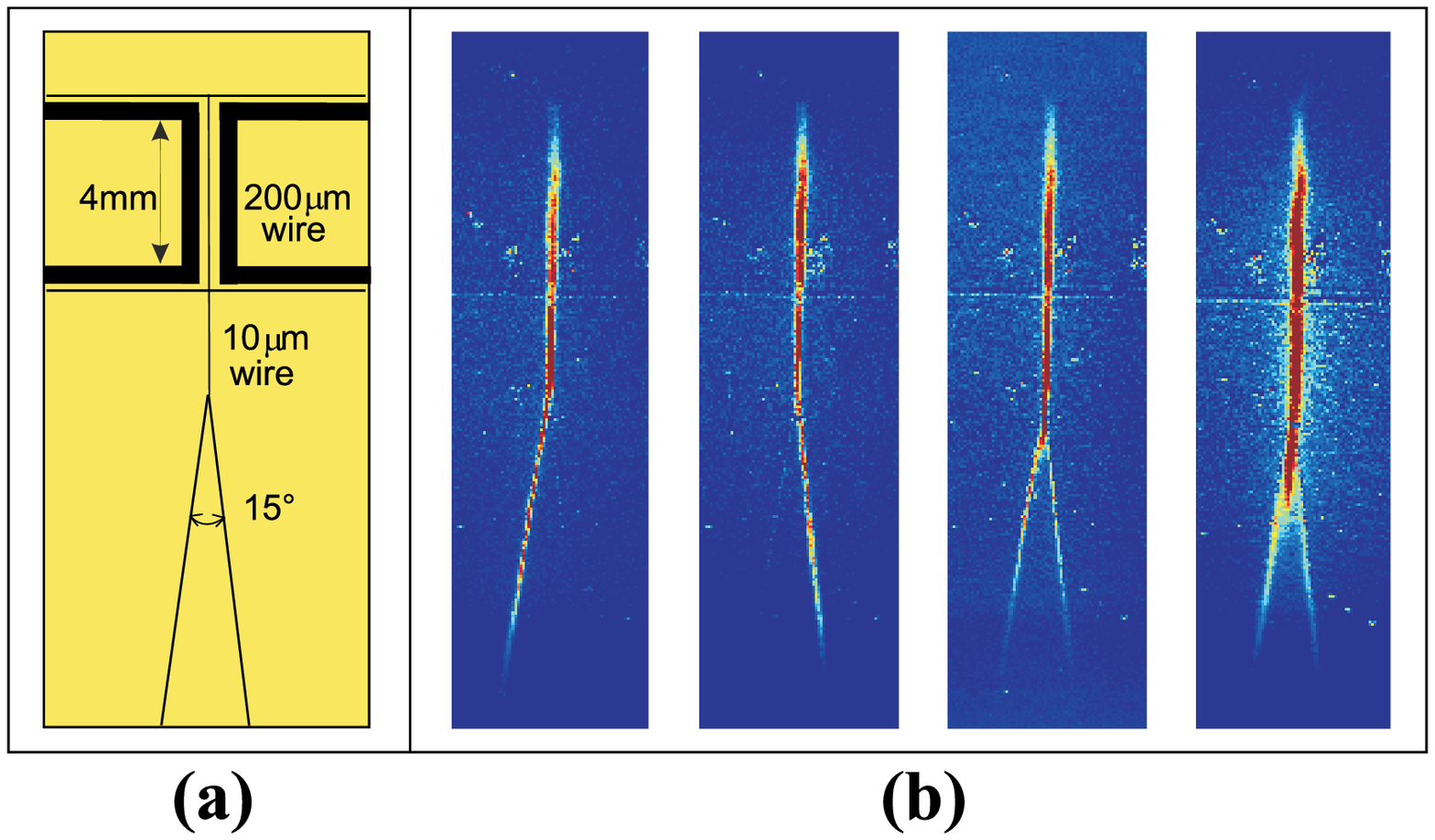}}\end{center}
    \vspace{2 cm}
    \caption{Beam splitter on a chip: (a) chip schematic and (b)
    fluorescence images of guided atoms. As explained
     in the text, the two large U-shaped 200$\mu m$
     wires are used to load atoms onto the 10$\mu m$ Y-shaped wire. In the first two pictures in
     (b), we drive current only through one side of the Y, therefore guiding atoms either to the
     left or to the right; in the next two pictures, taken at two different guide gradients, the current is
     divided in equal parts and the guided atoms split into both sides.}
    \label{chipBS}
\end{figure}

In this letter we describe experiments which join the above,
namely demonstrating a nanofabricated beam splitter for guided
atoms using microscopic magnetic guides on an {\em Atom Chip} (see
Fig.~\ref{chipBS}).

By bringing atoms close to electric and magnetic structures, one
can achieve high gradients to create microscopic potentials with a
size comparable to the de-Broglie wavelength of the atoms, in
analogy to mesoscopic quantum electronics
\cite{Schmiedmayer,MesoHinds}. It is possible to design quantum
wells, quantum wires and quantum dots for neutral atoms and
further combine these elements to form more complex structures. A
large variety of these microscopic potentials can be designed by
using the interaction $V = -\vec{\mu} \cdot \vec{B}$ between a
neutral atom with magnetic moment $\vec{\mu}$ and the magnetic
field $\vec{B}$ generated by current carrying structures
\cite{mesoscop,Vlad61,Wein95}. Mounting the wires on a surface,
allows elaborate designs with thin wires which can sustain sizable
currents \cite{Wein95}.  Such surface mounted atom optical
elements were recently demonstrated for large structures (wire
size $\approx 100 \mu m$) \cite{ChipGuide_large,fortagh,Haensch},
and nanofabricated structures \cite{ron}, the latter achieving the
scales required for mesoscopic physics and quantum information
proposals with microtraps \cite{Qcomp}.

The simplest configuration for a magnetic guide is a straight
current carrying wire \cite{Schm92,WireGuide,Vlad61}. The magnetic
field at a distance $r$ from the wire is given by $B =
\frac{\mu_{o}}{2 \pi} \frac{I}{r}$, where $I$ is the wire current.
Atoms in the high field seeking state are guided in Kepler orbits
around the wire (Kepler guide). By adding a homogeneous bias field
one can produce a 2-dimensional minimum of the potential at a
distance $\frac{\mu_{o}}{2 \pi} \frac{I}{B_{bias}}$ from the wire
\cite{Frisch33} and guide atoms in the low field seeking state
(side guide).

By combining two of these guides, it is possible to design
potentials where at some point two different paths are available
for the atom. This can be realized using different configurations,
among which the simplest and most advantageous is a Y-shaped wire
(Fig. 1a) \cite{Heiblum}. Such a beam splitter has one accessible
input for the atoms, that is the central wire of the Y, and two
accessible outputs corresponding to the right and left wires.
Depending on how the current $I$ in the input wire is sent through
the Y, atoms can be directed to the output arms of the Y with any
desired ratio (Fig.~\ref{chipBS}b).

The Y beam splitter can be created either as a Kepler guide or as
a side guide.  We previously performed preliminary experiments
studying such a beam splitting potential using free standing wires
\cite{mesoscop}. In the experiment reported here, we study a beam
splitter created by a Y-shaped wire on a nano-fabricated
\emph{Atom Chip}.

Our experiments are carried out using laser cooled  Li atoms. A
detailed description of the apparatus and the atom trapping
procedure is given in \cite{ron,denschlag}.

The \emph{Atom Chip} consists of a 2.5$\mu m$ thick gold layer
deposited onto a GaAs substrate. This gold layer is patterned
using standard nanofabrication techniques. A schematic of the
wires on the \emph{Atom Chip} used for this experiment is shown in
Fig. \ref{chipBS}a. It includes, besides the beam splitter, a
series of magnetic traps to transfer atoms into smaller and
smaller potentials: the large U-shaped wires are 200$\mu$m wide
and provide a quadrupole potential if combined with a homogeneous
bias field \cite{ron,Haensch,alb}, while the thin Y-shaped wire is
10$\mu m$ wide. An additional U-shaped 1mm thick wire is located
underneath the chip in order to assist with the loading of the
chip.

The atoms are loaded onto the \emph{Atom Chip} using our standard
procedure (see details in \cite{ron}): Typically $10^{8}$ cold
$^7Li$ atoms are accumulated in a "reflection MOT"
\cite{Raa87,LiMOT,pyramid} and transferred to the splitting
potential in the following steps: Atoms are first transferred into
the MOT generated by the quadrupole field of the U-wire (I=17A,
$B_{bias}=$6G) underneath the chip. Then, the laser light is
switched off, leaving the atoms confined only by the magnetic
quadrupole field of the U-wire.  Atoms are then further compressed
and transferred into a magnetic trap generated by the two 200$\mu
m$ wires on the chip (I=2A, $B_{bias}$=12G), compressed again and
transferred into the $10 \mu m$ guide (I=0.8A, $B_{bias}$=12G).
Each compression is achieved by decreasing the current generating
the larger trap to zero and simultaneously switching on the
current generating the smaller trap over a time of 10ms. Typically
we transfer $>10^{6}$ atoms into the 10$\mu m$ guide
\cite{loadingeff}, which has a typical transverse trap frequency
of $\omega=2\pi\times 6 kHz$.

The properties of the beam splitter are investigated by letting
the atoms propagate along the guide for some time due to their
longitudinal thermal velocity.  The resulting atom distribution is
measured by fluorescence images taken by a CCD camera looking at
the atom chip surface from above. For this a short ($< 0.5$ms)
molasses pulse is applied. The pictures shown in
Fig.~\ref{chipBS}b are such images taken after 16ms of guiding in
the beam splitter. The first two pictures are obtained at
$B_{bias}$=12G by sending 0.8A only through one of the output
wires; atoms can therefore turn either left or right. In the third
and fourth pictures the atoms experience a splitting potential,
the current being sent equally through both out-going arms of the
Y-shaped wire. The images are taken at bias field 12G and 8G
respectively. At 12G the atoms are clearly more compressed.

By changing the current ratio between the two outputs, and
simultaneously keeping the total current constant, it is possible
to control the probability of going left and right.  Typical data
for a beam splitter experiment using 8G bias field are shown in
Fig. \ref{ratios}. Here, the number of atoms in each arm is
determined by summing over the density distribution. When the
current is not balanced, the side carrying more current is
preferred due to the larger transverse size of the guiding
potential. It can be noted that the 50/50 atomic splitting ratio
occurs for a current ratio different from one half. This is due to
an additional 3G field directed along the input guide to make a
Ioffe-Pritchard configuration and prevent Majorana spin flips;
such a field introduces a difference in the output guides which
can be compensated with different currents. The solid lines shown
in Fig. \ref{ratios} are obtained with Monte Carlo simulations of
an atomic sample at T=250$\mu$K propagating in the Y beam
splitter.
%The temperature is an important parameter in order to achieve an
%agreement with the experiment: The calculations show that colder
%atoms are more sensitive to a small discrepancy from the symmetric
%case and therefore switch faster from left to right.

Before discussing the Y beam splitter in detail, one should note
some properties of the beam splitting potential created by the
Y-shaped wire and a homogeneous bias field as shown in Fig.
\ref{map}: (1) For the in-coming arm of the Y and for the two
out-going arms, far away from the splitting point, we have simple
side guide potentials. (2) The potential for the two out-going
guides is tighter than for the in-coming guide and its minimum is
at half distance from the chip surface. This is caused by the fact
that the in-coming guide is formed by a current which is twice
that of the out-going guides. It should also be noted that due to
the change in direction of the output wires, the bias field has
now a component along the guides which contributes to the
Ioffe-Pritchard field. (3) The splitting point of the potentials
is {\em not} at the geometrical splitting point of the wires. This
can be seen in the pictures of Fig.~\ref{chipBS}b. The actual
split point of the potential is located after the geometrical
split. Precisely, it occurs when the distance between the output
wires is given by $d_{split}=\frac{\mu_{o}}{2\pi}
\frac{I}{B_{bias}}$, which is equal to the height above the chip
of the input guide. (4) An additional potential minimum appears
between the geometric splitting point of the Y-wire and the
splitting point of the potential, forming a fourth port.

The different location of the potential split, and the additional
inaccessible fourth port of the beam splitting potential, can be
explained simply by taking two parallel wires with current in the
same direction and adding a homogeneous bias field along the plane
containing the wires and directed orthogonal to them. Depending on
the distance $d$ between the wires one observes three different
cases: (i) if $d<d_{split}$, two minima are created one above the
other on the axis between the wires. In the limit of $d$ going to
zero, the barrier potential between the two minima goes to
infinity (in approximation of infinitely thin wires) and the
minimum closer to the wires plane falls onto it. (ii) if
$d=d_{split}$, the two minima fuse into one. (iii) if
$d>d_{split}$ two minima are created one above each wire. The
barrier between them increases with the wire distance and we
eventually obtain two independent guides. In the Y beam splitter
one encounters all three cases moving along the beam splitter
axis. This is shown in detail in Fig.~\ref{map}b and c, which
present two projections, onto the beam splitter plane and
orthogonal to it respectively.

The dynamic of an atom propagating through the Y beam splitter
potential is best described by a scattering process in restricted
space from in-coming modes into out-going modes. As in most
scattering processes in free space, we expect some back scattering
into the in-coming mode. Additional back scattering mechanisms due
to the guides also occur: For instance, the output guides have
higher transverse gradients because each of them carries half of
the current. This gives rise to a reflection probability due to a
mismatch of modes. Another contribution comes from the direction
change of the input guide as it gets closer to the chip surface.
From the atomic distribution observed in the experiment we could
estimate a back-reflection of less than $20\%$ at the splitting
point. This may be further minimized by varying the potential
shape and choosing different geometries. In addition the fourth
port, caused by the second minimum before the split point, induces
a loss rate since atoms taking that route will hit the surface.

The Y configuration enables a 50/50 splitting over a wide range of
experimental parameters due to its {\em inherent symmetry}
relative to the incoming guide axis, where by inherent we mean
that the symmetry of the potential is maintained for different
magnitudes of current and bias field, and for different incoming
transverse modes. This was also numerically confirmed up to the
first 35 modes. The atom arriving at the splitting junction
encounters a symmetric scattering potential, and will thus have
equal right-left amplitudes regardless of the specific current and
bias field in use. Therefore, such a beam splitter should allow
inherently coherent splitting for multi-mode propagation. This
symmetric splitting may only be corrupted by breaking the symmetry
of the potential, for example by a rotation of the bias field
direction, or with a current imbalance.

This is an advantage over beam splitter designs for guided matter
waves which rely on tunneling \cite{andersson}. In such a
configuration, two side guides coming close together and going
apart again, the potential at the closest point exhibits two
guides separated by a potential barrier. Here the disadvantage is
that the splitting ratio for an incoming wave packet is vastly
different for different propagating modes, since it depends
strongly on the tunneling probability. A further disadvantage is
that, even for single mode splitting, the barrier width and height
and consequently the splitting amplitudes are extremely sensitive
to changes in the current and bias field.

The back scattering and the inaccessible fourth port of the Y beam
splitter may be, at least partially, overcome using different beam
splitter designs like the ones shown in Fig.~\ref{config}. The
configuration shown in Fig.~\ref{config}a has two wires which run
parallel until a given point and then go apart. In case the bias
field is chosen to exactly fulfill case (ii) of the above
discussion, the splitting point of the potential is ensured to be
that of the wires and the height of the potential minimum above
the chip surface is maintained throughout the device (in the limit
of small opening angle). Furthermore, no fourth port appears in
the splitting region. In Fig.~\ref{config}b we present a more
advanced design. Here a wave guide is realized with two parallel
wires with currents in opposite directions and a bias field
perpendicular to the chip surface. The splitting potential is
designed in order to have input and output guides with identical
characteristics, therefore eliminating reflections due to
different guide gradients. On the other hand, this multi-wire
configuration might be more difficult to integrate in a complex
network.

In conclusion, we have realized a beam splitter for guided atoms,
with a design that ensures symmetry under a wide range of
experimental parameters, and which we have shown can be further
developed to bypass the main drawbacks. This device could find
applications in atom interferometry and in the study of
decoherence processes close to a surface. Furthermore, this basic
element could be integrated into more complex quantum networks
which would form the base for advanced applications such as
quantum information processing.

We would like to thank  A. Chenet, A. Kasper, S. Schneider and A.
Mitterer for help in the experiments.  {\em Atom Chips} used in
the preparation of this work and in the actual experiments were
fabricated at the Institut f\"{u}r Festk\"{o}rperelektronik,
Technische Universit\"{a}t Wien, Austria, and the Sub-micron
center, Weizmann Inst. of Science, Israel. We thank E. Gornik, C.
Unterrainer and I. Bar-Joseph of these institutions for their
assistance. This work was supported by the Austrian Science
Foundation (FWF), project SFB F15-07, and by the ACQUIRE
collaboration (EU, contract Nr. IST-1999-11055). B.H. acknowledges
financial support from Svenska Institutet.

%\vspace{-0.5 cm}
%\include{BSref}

\newpage
\begin{figure}
    \begin{center}\hspace{0mm}\mbox{\input epsf
\epsfxsize0.9\columnwidth\epsfbox{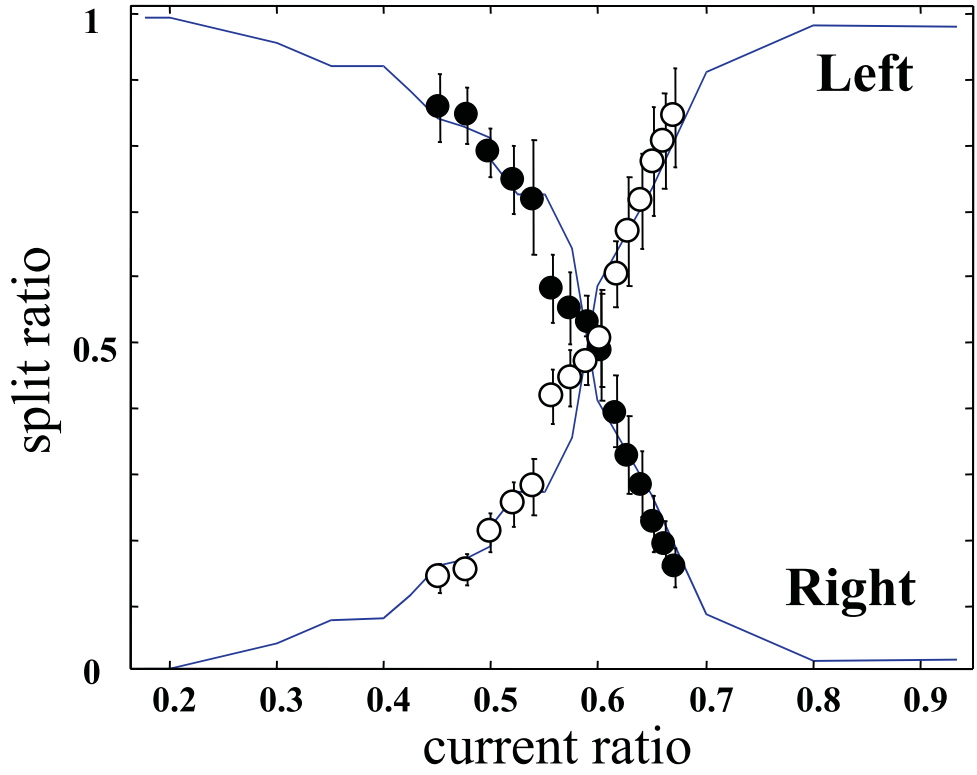}}\end{center}
    \vspace{2cm}
    \caption{Switching atoms between left and right by changing the current ratio
    in the two outputs and keeping the total current constant at 0.8A. The points are
    measured values while the lines are obtained from MC simulations. The kinks in the lines
    are due to MC statistics.}
    \lb{ratios}
\end{figure}

\newpage
\begin{figure}
 \begin{center}\hspace{0mm}\mbox{\input epsf
\epsfxsize0.9\columnwidth\epsfbox{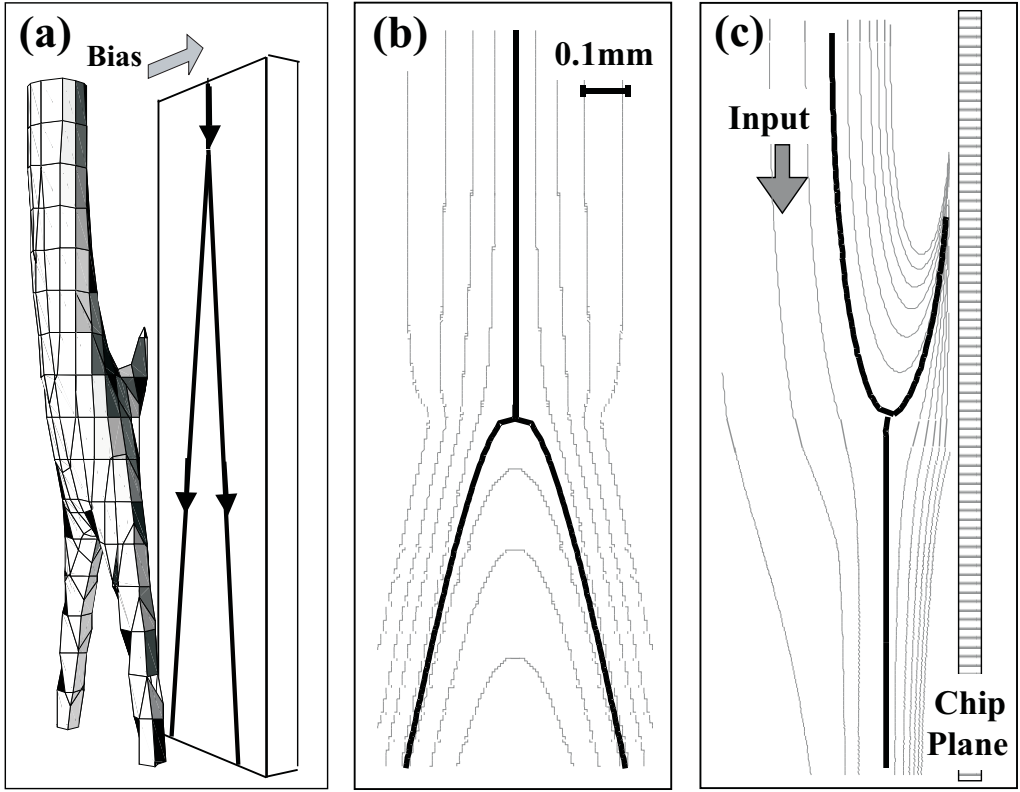}}\end{center}
 \vspace{2cm}
 \caption{Detailed properties of the Y splitting potential:
 (a) shows the 1.5G equipotential surface above the {\em Atom Chip} surface;
 (b) shows the position of the potential minima
 (black line) projected onto the
 surface; (c) shows the minima location above the surface. A second minimum
 closer to the chip surface occurs in the
 region between the wire splitting and the actual split point of the potential.
 In (b) and (c) the gray dotted lines are equipotential lines.
 These plots are generated at wire current 0.8A and
 $B_{bias}$=6G.}
 \lb{map}
\end{figure}

\newpage
\begin{figure}
 \begin{center}\hspace{0mm}\mbox{\input epsf
\epsfxsize0.9\columnwidth\epsfbox{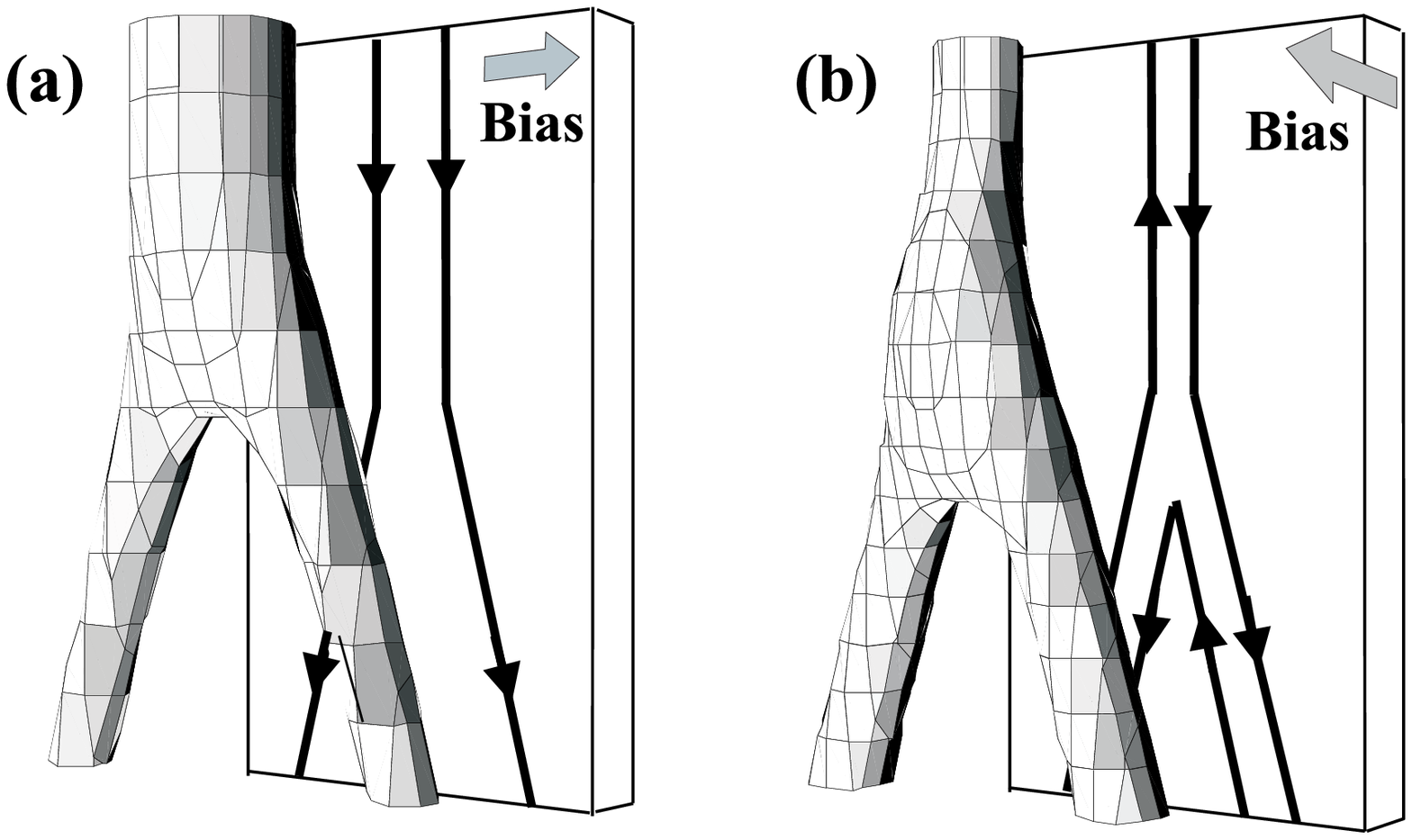}}\end{center}
 \vspace{2cm}
 \caption{Y beam splitter designs: The splitting potential in (a)
 is realized with two wires running parallel until a
 given point and then going apart; (b) is a more advanced Y beam
 splitter where the output guides have the same characteristics as
 the input guide, in order to minimize the back-scattered amplitude.}
 \lb{config}
\end{figure}


\begin{references}
%\vspace{-1.5 cm}
\bibitem{AtomOptics}   For an overview see: C.S. Adams, M. Sigel, J. Mlynek,
Phys. Rep. \textbf{240}, 143 (1994); \emph{Atom Interferometry},
edited by P. Berman (Academic Press, 1997). %and references therein.

\bibitem{ertmer}
K. Bongs {\em et al.},  Phys. Rev.
    Lett. {\bf 83}, 3577 (1999).

\bibitem{Schm92}
    J. Schmiedmayer in {\it XVIII International Conference on
    Quantum Electronics:} Technical Digest, edited by G.~Magerl
    (Technische Universit\"{a}t Wien, Vienna 1992), Series 1992, Vol.
    9, 284 (1992); Phys. Rev. A {\bf 52}, R13 (1995);
    Appl. Phys. B {\bf 60}, 169 (1995).

\bibitem{WireGuide}
    J. Denschlag, D. Cassettari, J. Schmiedmayer,
    Phys. Rev. Lett. \textbf{82}, 2014 (1999).

\bibitem{mesoscop}
    J. Denschlag, D. Cassettari, A. Chenet, S. Schneider, J. Schmiedmayer,
    Appl. Phys. B {\bf 69}, 291 (1999).

\bibitem{ChipGuide_large}
    D. M\"uller, {\em et al.},  Phys. Rev.
    Lett. {\bf 83}, 5194 (1999);\\
    N. H. Dekker, {\em et al.},  Phys. Rev.
    Lett. {\bf 84}, 1124 (2000).

\bibitem{Hindsguide}
    M. Key {\em et al.}, Phys. Rev.
    Lett. {\bf 84}, 1371 (2000).

\bibitem{ron}
    R. Folman, P. Kr\"uger, D. Cassettari, B. Hessmo, T. Maier, J. Schmiedmayer,
    quant-ph/9912106; Phys. Rev. Lett, in print.

\bibitem{HollowFibers}
    M. A. Ol'Shanii {\em et al.}, Opt. Comm. {\bf 98}, 77 (1993);
    S. Marksteiner {\em et al.}, Phys. Rev. A {\bf 50}, 2680 (1994);
    experiments: M.J. Renn {\em et al.}, Phys. Rev. Lett. {\bf 75}, 3253
    (1995); H. Ito {\em et al.}, Phys. Rev. Lett. {\bf 76}, 4500 (1996).

\bibitem{donught}
    S. Kuppens {\em et al.}, Phys. Rev. A {\bf 58}, 3068 (1998).

\bibitem{channeling}
    C. Salomon {\em et al.}, Phys. Rev. Lett. {\bf 59}, 1659
    (1987); V. I. Balykin {\em et al.}, Opt. Lett. {\bf 13} 958
    (1988); C. Keller, PhD Thesis, Universit\"at Wien (1999);
    C. Keller {\em et al.}, Appl. Phys. B {\bf 69}, 303
    (1999).

\bibitem{Schmiedmayer}
     J. Schmiedmayer, Eur.\ Phys.\ J.\ D {\bf 4}, 57 (1998).

\bibitem{MesoHinds}
    E. A. Hinds, I. G. Hughes, J. Phys. D: Appl. Phys. {\bf 32} 119
    (1999).

\bibitem{Vlad61}
    V. V. Vladimirskii, Sov. Phys. JETP {\bf 12}, 740 (1961).

\bibitem{Wein95}
    J. D. Weinstein, K. Libbrecht, Phys. Rev. A. {\bf 52}, 4004
    (1995);
    M. Drndic {\em et al.}, Appl. Phys. Lett. {\bf 72}, 2906
    (1998).


\bibitem{fortagh}
    J. Fortagh {\em et al.}, Phys. Rev. Lett. {\bf 81}, 5310 (1998).

\bibitem{Haensch}
    J. Reichel, W. Haensel, T. W. Haensch, Phys. Rev. Lett. {\bf 83}, 3398 (1999).



\bibitem{Qcomp}
    T. Calarco, D. Jaksch, E. A. Hinds, J. Schmiedmayer, J. I. Cirac, P. Zoller,
    Phys.\ Rev.\ A {\bf 61}, 022304 (2000).

\bibitem{Frisch33}
    R. Frisch, E. Segre, Z. f. Physik {\bf 75}, 610 (1933).

\bibitem{Heiblum}The Y configuration has been studied in quantum
electronics: T. Palm and L. Thyl\'en, Appl. Phys. Lett. {\bf 60},
237 (1992); J. J. Wesstr\"om, Phys. Rev. Lett. {\bf 82}, 2564
(1999).


%\bibitem{maier}
%    The chip is produced using standard nano-fabrication methods. A deta\'{\i}led account
%    will be given in: T. Maier {\em et al.}, to be published.

\bibitem{denschlag}
    J. Denschlag, PhD Thesis, Universit\"at Innsbruck (1998).

\bibitem{alb} A. Haase, D. Cassettari, B. Hessmo, J. Schmiedmayer,
submitted to Phys.\ Rev.\ A.

\bibitem{Raa87}
    E. L. Raab {\em et al.}, Phys. Rev. Lett. {\bf 59}, 2631 (1987).

\bibitem{LiMOT} The "reflection MOT" configuration has two laser beams
impinging onto the chip at 45 degrees and being reflected by the
gold surface. Since the wires are just defined by $10 \mu m$
etchings on the gold layer, the chip behaves as a mirror.
Eventually two counter-propagating beams are on the third axis,
parallel to the chip surface. The MOT quadrupole is produced by
coils with axis parallel to one of the 45 degrees laser beams.
After loading, the MOT atoms are cooled down to 200$\mu K$ by
shortly changing intensity and detuning of the laser beams.

\bibitem{pyramid}
    K. I. Lee {\em et al.}, Opt. Lett.
    {\bf 21}, 1177 (1996); see also \cite{Haensch}.

\bibitem{loadingeff}
The major atom loss occurs in the transfer between MOT and
magnetic trap, which has a typical efficiency of 5 - 10\% (with an
unpolarized atomic sample). In the following steps the transfer
efficiency, about 50\%, is mainly given by the strong compression
and therefore strong adiabatic heating of the atomic sample.

\bibitem{andersson} E. Andersson, M. T. Fontenelle, and S.
Stenholm, Phys. Rev. A \textbf{59}, 3841 (1999).


\end{references}
\end{document}